\documentstyle[pre,aps,multicol,epsf]{revtex}
\begin{document}
\title{Exchange Mediated Interaction of Dislocations and
Deformation Hardening of Invar Alloys}
\author{M. Molotskii and V. Fleurov}
\address{
School of Physics and Astronomy, Beverly and Raymond Sackler
Faculty of Exact Sciences.\\ Tel Aviv University, Tel Aviv 69978,
Israel}
\date{\today}
\maketitle
\begin{abstract}
We propose an explanation of anomalies observed in the behavior of
deformation hardening of Invars. An extremely strong volume
magnetostriction, typical of Invars, results in an enhancement of
the exchange contribution to the energy of the dislocation system
by about three orders of magnitude with respect to its value in
conventional ferromagnetics. Both the self energy of individual
dislocations and the interaction between them, determined usually
only by the elastic deformation, in Invars is strongly diminished
by the contribution of the exchange energy. This fact explains a
strong suppression of the deformation hardening rate in the second
stage as well as its strong temperature and content dependence in
Invars.
\end{abstract}
\pacs{81.40.C,E +75.50B +75.30.E}

\section{Introduction}

Quite a few very unusual physical properties characterize FCC
ferromagnetic Invar alloys. We can mention nearly zero thermal
expansion coefficient, giant values of the volume
magnetostriction, a strong pressure dependence of the Curie
temperature and of the saturation magnetization, anomalies in the
temperature and magnetic field dependence of the elastic constant
and other. The origin of the Invar magnetic and structural
instabilities has been intensively studied during the last decade
(see
\cite{w90,mj90,msw91,ehmsm93,aessj95,seaehr95,wsnsah97,invar97,saj99}
and references therein).

Since actually all physical properties of Invars are anomalous,
one should not wonder that their plastic properties also strongly
differ from those of the conventional FCC alloys
\cite{br69,ehy72,eh79,fgs80,rss85,r87,gs96}. We can mention here
the critical resolved shear stress in Invars which grows with
decreasing temperature with a rate, which is tens times larger
than that of the normal alloys. An explanation of this effect has
been recently proposed in our paper \cite{mf00a} where the
exchange interaction of the dangling d-states of the dislocation
cores and solute atoms (e.g., Ni atoms in FeNi Invar alloys) was
considered. The current paper addresses the problem of the
deformation hardening of Invars by considering the long range
interaction between dislocations which is mediated by the exchange
energy variation induced by the elastic strains in the vicinity if
the dislocations. It will be demonstrated below that the type of
interactions appears to be anomalously strong in Invar alloys.

A study of dislocation properties in ferromagnetic alloys started
more than half a century ago by Brown \cite{b40-51} who discussed
the influence of edge dislocations on the magnetization of
ferromagnetics in a strong magnetic field. This direction of study
was continued by Seeger and Kronm\"uller \cite{sk60,ks61} (see
also \cite{t69,k80,n87}). However, only linear magnetostriction
caused by the shear stresses in the vicinities of dislocations
were taken into account in those studies. As for the volume
magnetostriction, it was generally neglected. This neglect seems
to be well justified in conventional ferromagnetics (such as, say,
Fe or Ni), since their volume magnetostriction is two or three
orders of magnitude smaller than the linear one \cite{b51}.
Considering the Invar alloys one should keep in mind that their
volume magnetostriction is two orders of magnitude larger than
that of the conventional ferromagnetics \cite{b51,v74} and, hence,
the neglect can be hardly justified.

Ferromagnetics are characterized by a very large contribution of
the exchange interaction in the internal energy of the crystal.
This contribution is proportional to the squared magnetization
\cite{v74}. Dislocations induce hydrostatic strains in their
vicinities which result in variations of the magnetization, and
hence, of the exchange energy. In Invars, with their anomalously
strong volume magnetostriction, the variation of the exchange
energy appears to be anomalously strong. One should expect, as a
result, an anomalously strong contribution of the exchange effects
in the interaction between dislocations.

Introducing the notion of the exchange interaction between
dislocations and paramagnetic obstacles has allowed the authors of
this paper to explain the principal features of the electro- and
magnetoplastic effects \cite{mf95,mf97,mf00b}. Recent studies
demonstrate that sort of a "chemical" bonding between dislocation
and obstacles may play an important part in plasticity of alloys
\cite{aggkt96}.

The exchange effects appear to be exceptionally strong in Invars,
as has been demonstrated in our previous paper \cite{mf00a}, in
which the anomalously strong temperature dependence of critical
resolved shear stress of Invars has been explained. We plan to
calculate in the current paper the dislocation self energy and
their interaction energy accounting both for the elastic and
exchange contributions. It will be shown that anomalously strong
volume magnetostriction in Invars results in a large exchange
energy variation, which becomes comparable with the elastic
energy. This relates both to the energy associated with individual
dislocations and to the interaction between the dislocations. This
mechanism will allow us to explain the anomalous behavior of the
deformation hardening of Invars.

\section{Exchange Interaction Contribution to the Dislocation
Self Energy in Invars}

We calculate in this section the change of the total energy of a
crystal resulting from an introduction of an edge dislocation.
This change may be called the dislocation self energy. There are
two principal contributions to the dislocation self energy in a
ferromagnetic. First, there is an elastic energy, associated with
the strain field induced by the dislocation in its vicinity.
Second, there is an exchange energy, appearing due to the fact
that this strain changes direction and absolute value of the local
magnetizations {\bf M}. The calculation of the elastic energy can
be found in literature (see, e.g. \cite{n87,sty91}). Here we shall
concentrate on the calculation of the exchange energy of a
dislocation. In particular, it will be shown below that the
variation of the magnetization direction makes a contribution
negligible as compared to that of the absolute value variation.
The same can be said about the exchange energy variation due to
the linear magnetostriction.

The density of the exchange energy $w_{ex}$ in a ferromagnetic is
presented by many authors and contains two terms (see, e.g.,
\cite{v74})
\begin{equation}
w_{ex}=-\frac{\omega M^2}{2} + \frac{\omega d^2(\nabla M)^2}{z}.
\label{exch}
\end{equation}
The first term here depends only on the absolute value of the
magnetization $M$, whereas, the second, gradient, term accounts
for the space variations of $M$. Here
\begin{equation}
\omega=\frac{3k_BT_C}{np^2_{eff}\mu_B^2} \label{exch1}
\end{equation}
is the molecular field constant. $d$ is the interatomic spacing,
$z$ is the first coordination number, $T_C$ is the Curie
temperature. $p_{eff}$ is the effective number of the Bohr
magnetons per one atom, $n$ is the density of atoms, $k_B$ is the
Boltzmann constant, $\mu_B$ is the Bohr magneton.

An edge dislocation creates a hydrostatic pressure \cite{sty91}
$$p(\rho,\varphi)=-\frac{\mu
b}{3\pi}\frac{1+\nu}{1-\nu}\frac{\sin\varphi}{\rho}$$
slowly decaying with the distance $\rho$ from the dislocation
core. Here $\mu$ is the shear modulus, $\nu$ is the Poisson
coefficient, $b$ is the value of the Burgers vector, $\varphi$ is
the angle counted from the direction of the Burgers vector. The
pressure $p(\rho,\varphi)$ induces a variation of the
magnetization in the vicinity of the dislocation which in the
linear, in the local pressure, approximation, can be represented
as
$$\Delta M(\rho,\varphi)=\alpha \overline{M} p(\rho,\varphi).$$
Here $\overline{M}$ is the uniform magnetization of the
ferromagnetic in the absence of dislocations, $\alpha$ is a
proportionality constant known empirically for various
ferromagnetics (see, e.g., \cite{hm81} for Invars).

Now we can calculate the variation of the exchange energy of the
ferromagnetic caused by an edge dislocation. The variation of the
first term in equation (\ref{exch}), depending only on the
absolute value of the magnetization, reads
$$\Delta w^{m}_{ex} = -\frac{\omega \overline{M}^2}{2} [\alpha^2
p^2(\rho,\varphi) + 2 \alpha p(\rho,\varphi)].$$
Integrating this expression over a plane perpendicular to the
dislocation axis, one finds that the dislocation exchange energy
per the dislocation unit length is
\begin{equation}
W_{ex}^m = - \frac{\omega\overline{M}^2 \alpha^2 b^2 \mu^2}{18\pi}
\left( \frac{1+\nu}{1-\nu} \right)^2 \ln\frac{R}{r_0}.
\label{exch2}
\end{equation}
The calculation of the integral (\ref{exch2}) was carried out
under the same approximations as those used in the calculation of
the elastic energy of the dislocation in a continuous medium
\cite{sty91}. The integral is cut off at small distances, $\rho
> r_0 \sim d$, and at large distances, $\rho < R$, with $R$ being
of the order of the average distance between the dislocations.

The gradient part of the dislocation exchange energy (the second
term in (\ref{exch})) is calculated in a similar way,
\begin{equation}
W_{ex}^\nu = \frac{\omega\overline{M}^2 \alpha^2 b^2 \mu^2}{9 \pi
z} \left(\frac{1+\nu}{1-\nu}\right)^2 \label{exch3}
\end{equation}
The quantity $ln\frac{R}{r_0}$ is about 10 for typical
concentrations of dislocations ($\sim 10^7$ to $10^8$cm$^{-2}$),
while $z=12$ in FCC lattices. Therefore, the gradient term
(\ref{exch3}) is about 10$^{-2}$ of the dislocation exchange
energy (\ref{exch2}).

This difference of the two contribution follows mainly from the
fact that the variation of the absolute value of the
magnetization, $\Delta M(\rho,\varphi)$, decays only as
$\rho^{-1}$ with the distance from the dislocation core, whereas
the gradient, $\nabla M(\rho,\varphi)$ decays more rapidly as
$\rho^{-2}$. We should also consider the contribution of the
linear magnetostriction appearing due to the shear deformations
induced by the dislocations. The change of the local magnetization
is proportional the gradient of the angle $\vartheta$ defining the
direction of the atomic spins. It means that the dislocation self
energy due to the liner magnetostriction has generally the
structure similar to that of the gradient term in (\ref{exch}) and
is also small. We shall neglect these two small contributions in
what follows and omit the superscript $m$ in the exchange energy
(\ref{exch2}).

We need also the elastic energy per dislocation unit length which
in the same approximation \cite{n87,sty91} reads
\begin{equation}
W_{el}=\frac{\mu b^2}{4\pi(1-\nu)} \ln\frac{R}{r_0}. \label{elast}
\end{equation}
Now the total self energy of a dislocation (per its unit length)
containing both elastic and exchange contributions can be
represented as
\begin{equation}
W=W_{el} + W_{ex} = f(T) W_{el} \label{total}
\end{equation}
where
\begin{equation}
f(T)=1 - \frac{\omega \overline{M}^2(T)\alpha^2 E}{9} \left(
\frac{1+\nu}{1-\nu} \right). \label{factor1}
\end{equation}
Here the equality
$$E = 2\mu (1 + \nu)$$
connecting the Young modulus $E$ with the shear modulus $\mu$ has
been used.

Since the magnetization $\overline{M}(T)$ is a function of
temperature, the exchange contribution to the total energy $W$ is
also temperature dependent. As for the elastic contribution
$W_{el}$ its very weak temperature dependence can be neglected. In
order to estimate the temperature dependence of the total energy
we may use the mean field approximation in the theory of the
second order phase transitions, according which the magnetization
is
\begin{equation}
\overline{M}(T) = M_0\sqrt{1 - \frac{T}{T_C}} \label{magn}
\end{equation}
where $M_0$ is the spontaneous magnetization of the ferromagnetic
at zero temperature. Then the factor (\ref{factor1}) can be
represented as
\begin{equation}
f(T)=1 - \frac{\omega M_0^2\alpha^2 E}{9} \left(
\frac{1+\nu}{1-\nu} \right) \left( 1 - \frac{T}{T_C} \right).
\label{factor2}
\end{equation}

The simple formula (\ref{magn}) for the magnetization works rather
well in usual ferromagnetics in a wide range below the Curie
temperature. As for Invar alloys, it holds up to the temperatures
which are 20 to 30K below $T_C$ \cite{w90}. Closer to $T_C$ the
magnetization does not go to zero but becomes very small and falls
down slowly with the increasing temperature.

Now we demonstrate that only in Invars the contribution of the
exchange energy may be of importance. Really, the exchange
correction in the factor (\ref{factor1}) is very small in
conventional ferromagnetics and $f$ is close to one. For example,
Ni is characterized by the following parameters: $M_0=0.510$kG,
$E=3\times 10^{12}$dyn/cm$^2$, $\nu = 0.276$, $\omega = 13800$
\cite{b51}. As for the quantity $\alpha$ its value according to
\cite{hm81,ks60} is $-3\times 10^{-13}$ (dyn/cm$^2$)$^{-1}$. Then
one finds that the exchange correction in (\ref{factor1}) does not
exceed 2$\times 10^{-4}$ and, hence, can be neglected in normal
ferromagnetics.

The situation changes dramatically in Invar alloys in which the
quantity $\alpha$ is one to two orders of magnitude larger than in
normal ferromagnetics \cite{ks60}. For example, the
Fe$_{0.65}$Ni$_{0.35}$ Invar is characterized by $\alpha = -
1.1\times 10^{-11}$(dyn/cm$^2)^{-1}$ \cite{hm81} which is nearly
40 times larger than in Ni. As a result, the exchange energy
contribution, proportional to $\alpha^2$, becomes one thousand
times larger. The factor $f$ for the Fe$_{0.65}$Ni$_{0.35}$ alloy
can be estimated. First, the molecular field constant $\omega$ can
be calculated using equation (\ref{exch1}). It is know for this
alloy that $T_C=503$K \cite{w90}, $p_{eff}=1.851$ \cite{seaehr95},
$n=8.72\times 10^{22}$cm$^{-2}$ which results in $\omega=8110$.
Then using the values $M_0=1.4kG$, $E=1.4\times
10^{12}$dyn/cm$^2$, $\nu = 0.3$ \cite{k69} one finds that $f=0.35$
at low temperatures ($T\ll T_C$), meaning that the exchange
effects are able to decrease the total dislocation self energy
(\ref{total}) by a factor of three.

\section{Exchange Energy Contribution to the Interaction between
Dislocations in Invars}

We may distinguish two types of interaction between dislocations
connected with the magnetic structure of the ferromagnetic. The
direct magnetic interaction between local magnetizations in the
vicinities of parallel edge dislocations was calculated by Krey
\cite{k69}. This interaction appears to be three orders of
magnitude weaker than the elastic interaction even in Invar alloys
with their anomalously strong variations of the local
magnetization. This allows us to neglect in what follows the
direct magnetic interaction of dislocations.

However, a special attention should be paid to the second type of
interaction, i.e., the interaction between the dislocations caused
by the exchange effects, which can be much stronger than the
direct magnetic interaction. Considering the simplest case of two
parallel edge dislocations in the same sliding plane, when they
have either the same or the opposite mechanical signs, we can
calculate the exchange energy contribution to the interaction
between the dislocations.

Let $\bf a$ be the vector connecting two dislocations. Then the
local pressure induced by these two dislocation in a point $\bf r$
is
\begin{equation}
p({\bf r}) = p_1({\bf r}) + p_2({\bf r} - {\bf a})
\label{pressure}
\end{equation}
where $p_1$ and $p_2$ are the partial local pressures induced by
the first and the second dislocations, respectively. Now repeating
the calculations of the previous section with the pressure
distribution (\ref{pressure}) one arrives at the total change of
the exchange energy of the ferromagnetic caused by the two
dislocation. Then subtracting twice the self energy $W_{ex}$
(\ref{exch3}) of the individual dislocations one may get the
contribution of the exchange energy into the interaction between
the dislocations,
\begin{equation}
{\cal W}_{ex} = \mp \frac{\omega \overline{M}^2 \alpha^2
b^2\mu^2}{9\pi} \left( \frac{1+\nu}{1-\nu}\right)^2
\left(\frac{1}{2} + \ln\frac{R}{a}\right). \label{exch4}
\end{equation}
The signs correspond either to the same mechanical signs ($-$) or
to the opposite mechanical signs ($+$) of the dislocations. The
force due to the exchange interaction acting between the
dislocations per their unit lengths is
\begin{equation}
F_{ex}=- \frac{\partial {\cal W}_{ex}}{\partial a} = \mp
\frac{\omega \overline{M}^2 \alpha^2 b^2\mu^2}{9\pi a}\left(
\frac{1+\nu}{1-\nu}\right)^2. \label{force1}
\end{equation}

The exchange contribution to the interaction between the
dislocations has the sign opposite to that of the elastic
interaction. Contrary to the elastic interaction, the exchange
interaction results in an attraction of the dislocations with the
same mechanical signs and a repulsion of the dislocations with the
opposite signs.

The elastic force between two parallel edge dislocations
\cite{sty91} is
\begin{equation}
F_{el} = \pm \frac{\mu b^2}{2\pi(1-\nu)a}. \label{force2}
\end{equation}
Therefore, the total force acting between the two dislocation is
\begin{equation}\label{force3}
F(a)=F_{el}(a) + F_{ex}(a) = f(T) F_{el}(a)
\end{equation}
where $f$ is the same factor (\ref{factor1}) introduced in the
previous section. If the configuration of the two dislocations is
more complicated than that considered here, the expressions for
the elastic and exchange contributions to the interaction will
differ. However, we believe that the relation (\ref{force3}) will
generally hold. Hence, the discussion carried out in the previous
section relates also to the forces acting between the
dislocations. In conventional ferromagnetics the total force
(\ref{force3}) nearly coincides with the elastic force
(\ref{force2}) ($f\approx 1$). However, in Invars the interaction
due to the exchange energy becomes anomalously strong at low
temperatures and the factor $f$ may become essentially smaller
than one. Meaning that the total interaction between the
dislocations becomes essentially smaller than the elastic one.

\section{Deformation Hardening of Invar Alloys}

The analysis of the interaction between dislocation creates a
ground for discussing some specific features of deformation
hardening in Invar alloys. The deformation hardening in normal FCC
metals and alloys is well studied (see, e.g., review \cite{h83}).
The hardening can be subdivided into three stages. The hardening
rate $\theta_{II}$ in the second stage, usually called the rapid
stage of the deformation hardening, is approximately ten times
larger than the hardening rate in the first stage. That is why one
can easily distinguish between them. The quantity $\theta_{II}$
hardly depends either on the temperature or on the alloy content.

According to the experiments \cite{ehy72,fgs80} non of these
features hold in Invar alloys. The hardening rate $\theta_{II}$
rapidly falls down with decreasing temperature and strongly
depends on the alloy content. At low temperatures ($T\ll T_C$) its
value is two to three times smaller than the value $\mu/300$
typical for conventional FCC alloys. Therefore, the first and the
second stages are separated not well enough.

The result of the previous two sections allow us to understand
this drastic change of the behavior of the deformation hardening
in Invar alloys. The hardening rate $\theta_{II}$ is proportional
to the interaction between dislocations in different sliding
systems \cite{n87}. Therefore, accounting for the exchange contribution to
the interaction between the dislocations, the hardening rate becomes
\begin{equation}\label{hardening}
\theta_{II} = f(T) \theta_{II,\ el}
\end{equation}
where $\theta_{II,\ el}$ is the hardening rate caused by the
elastic interaction only.

As discussed above the factor $f(T)$ is nearly one in conventional
ferromagnetics, whereas in Invars this factor becomes at low
temperatures essentially smaller than one. $f(T)$ in Invars
decreases with the decreasing temperature which should lead to a
corresponding decrease of the deformation hardening rate
$\theta_{II}$. Such a temperature dependence was really observed
in \cite{fgs80}. In case of the Fe$_{0.65}$Ni$_{0.35}$ the factor
$f(T)$ can become as small as $f=0.35$ at $T\ll T_c$, meaning that
the hardening rate $\theta_{II}$ in the Invar should be
approximately three times smaller that its pure "elastic" value
$\theta_{II,\ el}$.

The exchange contribution to the dislocation interaction becomes
negligible at temperatures above $T_C$, meaning that $\theta_{II}$
measured at high temperatures is purely elastic one, $\theta_{II,\
el}$. This allows us to use the experimental data (above and below
$T_C$) presented in references \cite{ehy72,fgs80} and estimate the
ratio $\displaystyle \frac{\theta_{II,\ Invar}} {\theta_{II,\
normal}}$ as lying in the range from 1/3 to 1/2. It agrees very
well with our theoretical estimate of the $f$ value.

The parameter $\alpha$ depends on the Invar alloy content. Even a
slightest deviation from the typical content
Fe$_{0.65}$Ni$_{0.35}$ causes a strong drop of the $\alpha$ value
and, hence, much stronger drop of the exchange effect
contribution. Just, for example, changing the Ni concentration in
this alloy from 35\% to 38\% results in a decrease of $\alpha$ by
a factor of two \cite{ks60}. Then, according to (\ref{factor2}),
the coefficient $f$ changes from 0.35 to 0.84. The exchange
effects, as a result, are essentially suppressed and the
deformation hardening rate in such an alloy is much closer to that
of a normal alloy. The variation of the coefficient $f$ with the
alloy content may provide an explanation of a strong dependence of
the deformation hardening rate on the Invar alloy content,
observed in references \cite{ehy72,fgs80}.

In order to demonstrate it, we calculate the dependence of the
deformation hardening rate $\theta_{II}$ on the concentration
dependence in FeNi Invar alloys (see figure 1). Unfortunately the
available experimental information is very limited. The value of
the coefficient $\alpha$ at room temperature is known only for the
Fe$_{0.65}$Ni$_{0.35}$ alloy \cite{hm81}. Reference \cite{ks60}
provides the $\alpha$ values for the Ni concentrations, 34.7\% and
44.9\%, but only at low temperatures, 4.2 and 20.4 K. To the best
of our knowledge there are no other data on the concentration
dependence of this coefficient. That is why we assume the
concentration dependence of $\alpha$ measured in \cite{ks60} at
low temperatures and scale it to the room temperature value of
$\alpha$ \cite{hm81}. This rather rough approximation can be
readily improved when new data on the $\alpha$ concentration
dependence will appear.

\begin{figure}[htb]
\epsfysize=15 \baselineskip \centerline{\hbox{
\epsffile{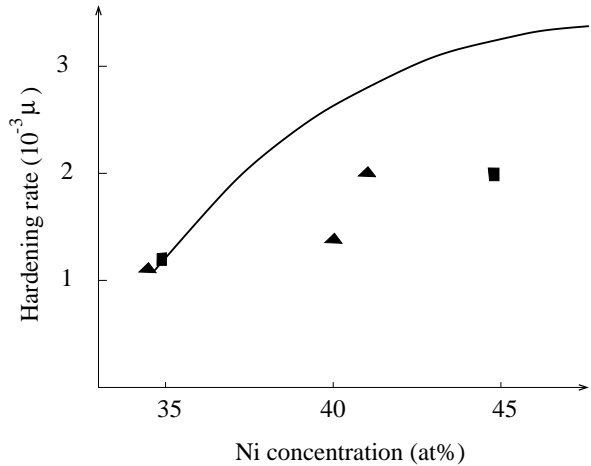}}}\vspace{.5cm} \caption{Dependence of the
deformation hardening rate of FeNi alloys on the Ni concentration.
Solid line --- theoretical calculation, available experimental
data are represented by black triangle \protect\cite{ehy72} and
black squares \protect\cite{fgs80}}
\end{figure}

The choice of the remaining parameters is straightforward. The
deformation hardening rate, $\theta_{II,\ el}$, due to elastic
strains is $\mu/300$ \cite{ehy72}. The elastic constants
$E=1.4\times 10^{12}$dyn/cm$^2$ and $\nu=0.3$ do not depend on the
alloy content at room temperature \cite{k69,hw73}. The Curie
temperature for Invar alloy is very high, so we may assume that
the magnetization has arrived at saturation at room temperature,
i.e., $\bar M = np_{eff}\mu_B$ \cite{w90}. Then using equations
(\ref{exch1}) and (\ref{factor1}) one gets
\begin{equation}\label{factor3}
  f=1 - \frac{1}{3}k_BT_c\alpha^2E\frac{1+\nu}{1-\nu}
\end{equation}

The dependence of the Curie temperature on the alloy content is
available in \cite{hw73}. Then using equation (\ref{hardening}) we
are able to calculate the concentration dependence of the
deformation hardening $\theta_{II}$ for FeNi alloy presented by
the solid line in figure 1. The theory correctly predicts an
increase of the deformation hardening with the increasing Ni
concentration. It is also in good quantitative agreement with the
experiment for the Fe$_{0.65}$Ni$_{0.35}$ alloy for which we have
reliable experimental value of the coefficient $\alpha$. We
observe a rather strong deviation of the theory for the other Ni
contents where we have been forced to use less reliable values of
$\alpha$, which may be the cause of this deviation. This
calculation demonstrates a necessity of additional measurements of
various characteristics of Invar alloys with varying contents in
order to verify the model proposed in this publication.

\section{Conclusions}

This study demonstrates that the peculiar behavior of the
deformation hardening of Invars cannot be explained by applying
the conventional approach which considers only the elastic strain
mediated interaction of dislocation. It is important to
incorporate also the influence of the exchange effect on the
interaction between dislocations which is anomalously strong in
Invars. The exchange mediated interaction between dislocations has
the sign opposite to the elastic contribution and may essentially
diminish the total interaction at temperatures well below the
Curie temperature. This decrease of the interaction is responsible
for the lower rate of the deformation hardening in the second
stage. This mechanism allows one also to explain the strong
temperature and content dependence of the deformation hardening
rate observed experimentally. An interesting case when the
exchange interaction is so strong that it changes the sign of the
total interaction needs a special attention and will be studied
elsewhere.

A more detailed experimental study of the temperature and content
dependence of the deformation hardening rate in Invars is
necessary in order to verify the theoretical model developed in
this paper. Equations (\ref{factor1}) and (\ref{hardening})
indicate that the temperature dependence of the deformation
hardening rate $\theta_{II}(T)$ is directly connected to
temperature dependence of the Invar magnetization
$\overline{M}(T)$. An experimental observation of such a
connection would provide a strong support for our model.


\begin{references}
\bibitem{w90} E.P.Wasserman, in {\em Ferromagnetic Materials}, ed.
K.H.J.Buschow and E.P.Wohlfarth (North-Holland, Amsterdam, 1990).
Vol. {\bf 5}, pp.237-321

\bibitem{mj90} E.G.Moroni and T.Jarlborg, Phys.Rev., {\bf B41},
9600 (1990)

\bibitem{msw91} E.G.Moroni, K.Schwarz, and D.Wagner, Phys.Rev., {\bf B43},
3318 (1991)

\bibitem{ehmsm93} P.Entel, E.Hoffmann, P.Mohn, K.Schwarz, and V.L.Moruzzi,
Phys.Rev., {\bf B47}, 8706 (1993)

\bibitem{aessj95} I.A.Abrikosov, O.Eriksson, P.Sonderlind, H.L.Silver, and
B.Johansson, Phys.Rev., {\bf B51}, 1058 (1995)

\bibitem{seaehr95} M.Schr\"oter, H.Ebert, H.Akai, P.Entel, E.Hoffmann, and G.G.Reddy,
Phys.Rev., {\bf B52}, 188 (1995)

\bibitem{wsnsah97} Y.Wang, G.M.Stocks, D.M.C.Nickolson, W.A.Shel\-ton, A.V.Antro\-pov,
and B.N.Har\-mon, J. Appl. Phys., {\bf 81}, 3873 (1997)

\bibitem{invar97} {\em The Invar Effect}: Centennial Symposium, ed.
J.Wittenauer (Minerals, Metals and Materials Society, Warrendale,
Pennsylvania, 1997)

\bibitem{saj99} M.van Schilfgaarde, I.A.Abrikosov, and B.Johansson, Nature, {\bf
400}, 46 (1999)

\bibitem{br69} C.F.Bolling and R.H.Richman, Philos.Mag., {\bf 19},
247, (1969)

\bibitem{ehy72} J.Echigoya, S.Hayashi, and Y.Yamamoto, Phys.Stat.Sol.,
(a), {\bf 14}, 463 (1972)

\bibitem{eh79} J.Echigoya and S.Hayashi, Phys.Stat.Sol., (a), {\bf 55}, 279
(1979)

\bibitem{fgs80} H.Flor, H.J.Gudladt, and Ch.Schwink, Acta Metal., {\bf 28},
1611 (1980)

\bibitem{rss85} I.Retat, Th.Steffens, and Ch.Schwink, Phys.Stat.Sol;, (a),
{\bf 92}, 507 (1985)

\bibitem{r87} I.Retat, Phys.Stat.Sol., (a), {\bf 99}, 121 (1981)

\bibitem{gs96} A.Gulayev and E.L.Svistunova, Scr.Mater., {\bf 35}, 501
(1996)

\bibitem{mf00a} M.Molotskii and V.Fleurov, Phys.Rev. {\bf B}
submitted; preprint - cond-mat/0011325

\bibitem{b40-51} W.F.Brown, Jr, Phys.Rev., {\bf 58}, 736 (1940);
{\bf 60}, 139 (1941); {\bf 82}, 94 (1951)

\bibitem{sk60} A.Seeger and H.Kronm\"uller, J.Phys.Chem.Sol., {\bf
12}, 298 (1960)

\bibitem{ks61} H.Kronm\"uller and A.Seeger, J.Phys.Chem.Sol., {\bf
18}, 93 (1961)

\bibitem{t69} H.Tr\"auble, in {\em Magnetism and Metallurgy}, ed.
A.E.Bercowitz and E.Kneller (Academic Press, New York, 1969), Vol.
2, pp. 621 - 687

\bibitem{k80} M.Kleman, In: {\em Dislocations in Solids}. ed. by
F.R.N.Nabarro (North-Holland, Amsterdam, 1980), vol. 5, pp.
349-402

\bibitem{n87} F.R.N.Nabarro, {\em Theory of Crystal Dislocations},
(Dover Publ., New York, 1987)

\bibitem{b51} R.M.Bozorth, {\em Ferromagnetism}, (D. van Nostrand C0, Toronto, 1951)

\bibitem{v74} S.V.Vonsovskii, {\em Magnetism}, (J.Wiley \& Sons, New York,
1974), Vol. 2

\bibitem{mf95} M.Molotskii and V.Fleurov, Phys.Rev., B{\bf 52}, 15829 (1995)

\bibitem{mf97} M.Molotskii and V.Fleurov, Phys.Rev.Lett., {\bf 78}, 2779
(1997)

\bibitem{mf00b} M.Molotskii and V.Fleurov, J.Phys.Chem., {\bf B104},
3812 (2000)

\bibitem{aggkt96} A.O.Anokhin, M.L.Galperin, Yu.N.Gornostyrev,
M.I.Katsnelson, and A.V.Trefilov, {\em Philos.Mag.}, {\bf B 73},
845 (1996)

\bibitem{sty91} T.Suzuki, S.Takeuchi, and H.Yoshinag, {\em Dislocation Dynamics and
Plasticity}, (Springer, Berlin, 1991)

\bibitem{hm81} H.Hayashi and N.Mori, Sol.St.Communs., {\bf 38}, 1057 (1981)

\bibitem{ks60} E.I.Kondorskii and V.L.Sedov, Zh.Eksp.Teor.Fiz., {\bf 38},
773 (1960) [Sov.Phys.JETP, {\bf 11}, 561 (1960)]

\bibitem{k69} U.Krey, Philos.Mag., {\bf 20}, 1295 (1969)

\bibitem{h83} P.Haasen, in {\em Physical Metallurgy,} ed. R.W.Cahn and
P.Haasen (North-Holland, Amsterdam, 1983), Part II, pp. 1341-1409

\bibitem{hw73} G.Hausch and H.Warlimont, Acta Metall., {\bf 21},
401 (1973)

\end{references}
\end{document}